\documentclass[12pt,3p,authoryear]{elsarticle}

\usepackage[hyphens]{url}
\usepackage{graphicx}
\usepackage{booktabs}
\usepackage[T1]{fontenc}
\usepackage{lmodern}
\usepackage{caption}
\usepackage{subfig}
\usepackage{amssymb, amsmath}
\usepackage[inline]{enumitem}
\usepackage{float}
\usepackage{tabularx}
\usepackage[dvipsnames, table]{xcolor}
\usepackage{ifxetex, ifluatex}
\usepackage{fixltx2e}
\usepackage[unicode=true, colorlinks]{hyperref}
\usepackage{cleveref}
\usepackage{tabu}
\usepackage{mathpazo}

\usepackage{easyReview}

\IfFileExists{upquote.sty}{\usepackage{upquote}}{}

\ifnum 0\ifxetex 1\fi\ifluatex 1\fi=0 
  \usepackage[utf8]{inputenc}

\else 
  \usepackage{fontspec}
  \ifxetex
    \usepackage{xltxtra,xunicode}
  \fi
  \defaultfontfeatures{Mapping=tex-text,Scale=MatchLowercase}

\fi

\IfFileExists{microtype.sty}{\usepackage{microtype}}{}

\usepackage{longtable}

\usepackage{lscape}

\usepackage{setspace}
\usepackage{lmodern}

\providecommand{\tightlist}{%
  \setlength{\itemsep}{0pt}\setlength{\parskip}{0pt}}
  
\AtBeginDocument{%
  \newcommand\myshade{80}
  \colorlet{mylinkcolor}{violet!\myshade!black}
  \colorlet{mycitecolor}{YellowOrange!\myshade!black}
  \colorlet{myurlcolor}{Aquamarine!\myshade!black}

  \hypersetup{
    breaklinks = true,
    bookmarks  = true,
    pdfauthor  = {},
    pdftitle   = {Comparison of Multi-response Estimation Methods},
    linkcolor  = mylinkcolor,
    citecolor  = mycitecolor,
    urlcolor   = myurlcolor,
    colorlinks = true,
  }
}
\biboptions{numbers,sort&compress}

\makeatletter
\providecommand{\doi}[1]{%
  \begingroup
    \let\bibinfo\@secondoftwo
    \urlstyle{rm}%
    \href{http://dx.doi.org/#1}{%
      doi:\discretionary{}{}{}%
      \nolinkurl{#1}%
    }%
  \endgroup
}
\makeatother

\begin{document}
\begin{frontmatter}

  \title{Comparison of Multi-response Estimation Methods}
  
    \author[KBM]{Raju Rimal\corref{c1}}
   \ead{raju.rimal@nmbu.no} 
   \cortext[c1]{Corresponding Author}
    \author[KBM]{Trygve Almøy}
   \ead{trygve.almoy@nmbu.no} 
  
    \author[KBM]{Solve Sæbø}
   \ead{solve.sabo@nmbu.no} 
  
      \address[KBM]{Faculty of Chemistry and Bioinformatics, Norwegian University of Life Sciences, Ås, Norway}
  
  \begin{abstract}
  Prediction performance does not always reflect the estimation behaviour of a method. High error in estimation may necessarily not result in high prediction error, but can lead to an unreliable prediction if test data lie in a slightly different subspace than the training data. In addition, high estimation error often leads to unstable estimates, and consequently, the estimated effect of predictors on the response can not have a valid interpretation. Many research fields show more interest in the effect of predictor variables than actual prediction performance. This study compares some newly-developed (envelope) and well-established (PCR, PLS) estimation methods using simulated data with specifically designed properties such as Multicollinearity in the predictor variables, the correlation between multiple responses and the position of principal components corresponding to predictors that are relevant for the response. This study aims to give some insights into these methods and help the researchers to understand and use them for further study. Here we have, not surprisingly, found that no single method is superior to others, but each has its strength for some specific nature of data. In addition, the newly developed envelope method has shown impressive results in finding relevant information from data using significantly fewer components than the other methods.
  \end{abstract}

\end{frontmatter}

\hypertarget{introduction}{%
\section{Introduction}\label{introduction}}

Estimation of parameters in linear regression models is an integral part of many research studies. Research fields such as social science, econometrics, chemometrics, psychology and medicine show more interest in measuring the impact of certain indicators or variable than performing prediction. Such studies have a large influence on people's perception and also help in policy-making and decisions. A transparent, valid and robust research is critical to improving the trust in the findings of modern data science research \citep{eu2019auethics}. This makes the assessment of measurement error, inference and prediction even more essential.

Technology has facilitated researchers to collect large amounts of data, however, often such data either contains irrelevant information or are highly redundant. Researchers are devising new estimators to extract information and identify their inter-relationship. Some estimators are robust towards fixing the multicollinearity (redundancy) problem, while others are targeted to model only the relevant information contained in the response variable.

This study extends \citep{rimal2019pred} with a similar multi-response, linear regression model setting and compares some well-established estimators such as Principal Components Regression (PCR), Partial Least Squares (PLSR) Regression, together with two new methods based on envelope estimation: Envelope estimation in predictor space (Xenv) \citep{cook2010envelope} and simultaneous estimation of the envelope (Senv) \citep{cook2015simultaneous}. The estimation processes of these methods are discussed in the \protect\hyperlink{estimation-methods}{Estimation Methods} section. The comparison is aimed at the estimation performance of these methods using multi-response simulated data from a linear model with controlled properties. The properties include the number of predictors, level of multicollinearity, the correlation between different response variables and the position of relevant predictor components. These properties are explained in the \protect\hyperlink{experimental-design}{Experimental Design} section together with the strategy behind the simulation and data model.

\hypertarget{simulation-model}{%
\section{Simulation Model}\label{simulation-model}}

\begin{figure}

{\centering \includegraphics[width=0.8\linewidth]{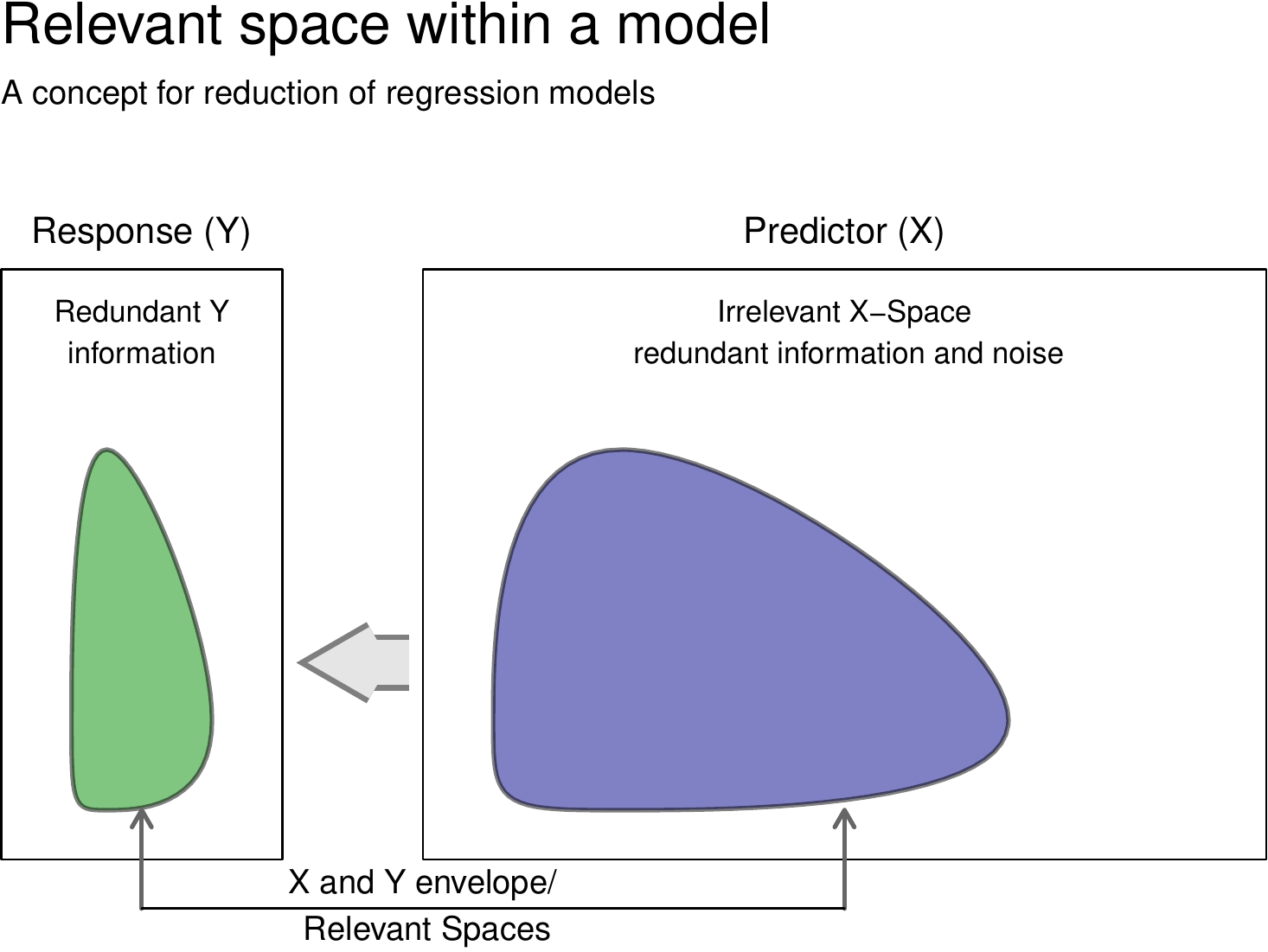} 

}

\caption{Relevant space in a regression model}\label{fig:relevant-space}
\end{figure}

As a follow-up, this study will continue using the same simulation model as used by \citet{rimal2019pred}. The data are simulated from a multivariate normal distribution where we assume that the variation in a response vector-variable \(\mathbf{y}\) is partly explained by the predictor vector-variable \(\mathbf{x}\). However, in many situations, only a subspace of the predictor space is relevant for the variation in the response \(\mathbf{y}\). This space can be referred to as the relevant space of \(\mathbf{x}\) and the rest as irrelevant space. In a similar way, for a certain model, we can assume that a subspace in the response space exists and contains the information that the relevant space in predictor can explain (Figure \ref{fig:relevant-space}).

Following the concept of relevant space, a subset of predictor components can be imagined to span the predictor space. These components can be regarded as relevant predictor components. \citet{Naes1985} introduced the concept of relevant components, which was explored further by \citet{helland1990partial}, \citet{naes1993relevant}, \citet{Helland1994b} and \citet{Helland2000}. The corresponding eigenvectors were referred to as relevant eigenvectors. A similar logic is introduced by \citet{cook2010envelope} and later by \citet{cook2013envelopes} as an envelope, as space spanned by the relevant eigenvectors \citep[p.101]{cook2018envelope}. See \citet{Rimal2018}, \citet{saebo2015simrel} and \citet{rimal2019pred} for in-depth background on the model.

\hypertarget{estimation-methods}{%
\section{Estimation Methods}\label{estimation-methods}}

Consider a joint distribution of \(\mathbf{y}\) and \(\mathbf{x}\) with corresponding mean vectors \(\boldsymbol{\mu}_y\) and \(\boldsymbol{\mu}_x\) as,

\begin{equation}
  \begin{bmatrix}
    \mathbf{y} \\ \mathbf{x}
  \end{bmatrix} 
  \sim \mathsf{N}
  \begin{pmatrix}
    \begin{bmatrix}
      \boldsymbol{\mu}_y \\
      \boldsymbol{\mu}_x \\
    \end{bmatrix}, &&
    \begin{bmatrix}
      \boldsymbol{\Sigma}_{yy} & \boldsymbol{\Sigma}_{yx} \\
      \boldsymbol{\Sigma}_{xy} & \boldsymbol{\Sigma}_{xx} 
    \end{bmatrix}
  \end{pmatrix}
  \label{eq:model-1}
\end{equation}

Here, \(\boldsymbol{\Sigma}_{xx}\) and \(\boldsymbol{\Sigma}_{yy}\) are variance-covariance of \(\mathbf{x}\) and \(\mathbf{y}\) respectively and \(\boldsymbol{\Sigma}_{xy}=\boldsymbol{\Sigma}_{yx}^t\) is the covariance matrix of \(\mathbf{x}\) and \(\mathbf{y}\). Let \(\mathbf{S}_{xx}\), \(\mathbf{S}_{yy}\) and \(\mathbf{S}_{xy}=\mathbf{S}_{yx}^t\) be the respective estimates of these matrices. A linear regression model based on \eqref{eq:model-1} is

\begin{equation}
\mathbf{y} = \boldsymbol{\mu}_y + \boldsymbol{\beta}^t\left( \mathbf{x} - \boldsymbol{\mu}_{x} \right) + \boldsymbol{\varepsilon}
\label{eq:reg-model}
\end{equation}

where \(\boldsymbol{\beta}=\boldsymbol{\Sigma}_{xx}^{-1}\boldsymbol{\Sigma}_{xy}\)
is the regression coefficients that define the relationship between \(\mathbf{x}\)
and \(\mathbf{y}\). With \(n\) samples, the least-squares estimate of
\(\boldsymbol{\beta}\) can be written as
\(\boldsymbol{\hat{\beta}}=\mathbf{S}_{xx}^{-1}\mathbf{S}_{xy}\). Here, as in many
situations, the estimator \(\mathbf{S}_{xx}\) for \(\boldsymbol{\Sigma_{xx}}\) can either be non-invertible or have small eigenvalues. In addition, \(\mathbf{S}_{xy}\), the estimator of \(\boldsymbol{\Sigma_{xy}}\), is often influenced by a high level of noise in the data. In order to solve these problems, various methods have adopted the concept of relevant space to identify the relevant components through the reduction of the dimension in either \(\mathbf{x}\) or \(\mathbf{y}\) or both. Some of the methods we have used for comparison are discussed below.

\emph{Principal Components Regression} (PCR) uses \(k\) eigenvectors of \(\mathbf{S}_{xx}\) as the number of components to span the reduced relevant space. Since PCR is based on capturing the maximum variation in predictors for every component it has added to the model, this method does not consider the response structure in the model reduction \citep{Jolliffe2002}. In addition, if the relevant components are not corresponding to the largest eigenvalues, the method requires a larger number of components to make precise prediction \citep{Alm_y_1996}.

\emph{Partial Least Squares} (PLS) regression aims to maximize the covariance between the predictor and response components (scores) \citep{DeJong1993}. Broadly speaking, PLS can be divided into PLS1 and PLS2 where the former tries to model the response variables individually, whereas the latter uses all the response variable together while modelling. Among the three widely used algorithms NIPALS \citep{wold75nipals}, SIMPLS \citep{DeJong1993} and KernelPLS \citep{Lindgren_1993}, we will be using KernelPLS for this study, which gives equivalent results to the classical NIPALS algorithm and is default in R-package \texttt{pls} \citep{mevik07_thepl}.

\emph{Envelopes} was first introduced by \citep{Cook2007a} as the smallest subspace that includes the span of true regression coefficients. The \emph{Predictor Envelope} (Xenv) identifies the envelope as a smallest subspace in the predictor space, by separating the predictor covariance \(\boldsymbol{\Sigma}_{xx}\) into relevant (material) and irrelevant (immaterial) parts, such that the response \(\mathbf{y}\) is uncorrelated with the irrelevant part given the relevant one. In addition, relevant and irrelevant parts are also uncorrelated. Such separation of the covariance matrix is made using the data through the optimization of an objective function. Further, the regression coefficients are estimated using only the relevant part. \citet{cook2010envelope}, \citet{cook2013envelopes} and \citet{cook2018envelope} have extensively discussed the foundation and various mathematical constructs together with properties related to the Predictor Envelope.

\emph{Simultaneous Predictor-Response Envelope} (Senv) implements the envelope in both the response and the predictor space. It separates the material and immaterial part in the response space and the predictor space such that the material part of the response does not correlate with the immaterial part of the predictor and the immaterial part of the response does not correlate with the material part of the predictor. The regression coefficients are computed using only the material part of the response and predictor spaces. The number of components specified in both of these methods during the fit influences the separation of these spaces. If the number of response components equals the number of responses, simultaneous envelope reduces to the predictor envelope, and if the number of predictor components equals the number of predictors, the result will be equivalent to ordinary least squares. \citet{cook2015simultaneous} and \citet{cook2018envelope} have discussed the method in detail. Further, \citet{helland2016algorithms} have discussed how the population models of PCR, PLS and Xenv are equivalent.

\hypertarget{experimental-design}{%
\section{Experimental Design}\label{experimental-design}}

An R \citep{coreR2018} package \texttt{simrel} \citep{Rimal2018, saebo2015simrel} is used to simulate the data for comparison. In the simulation, the number of observations is fixed at \(n = 100\), and the following four simulation parameters are varied to obtain data with a wide range of properties.

\begin{description}
\tightlist
\item[\textbf{Number of predictors: (\texttt{p})}]
In order to cover both tall \((n>p)\) and wide \((p>n)\) cases, \(p= 20\) and \(p= 250\) number of predictors are simulated.
\item[\textbf{Multicollinearity in predictor variables: (\texttt{gamma})}]
A parameter \texttt{gamma} \((\gamma)\) controls the exponential decline of eigenvalues in \(\boldsymbol{\Sigma_{xx}} (\lambda_i, i = 1, \ldots p)\) as,
\begin{equation}
  \lambda_i = e^{-\gamma(i-1)}, \gamma > 0 \text{ and } i = 1, 2, \ldots p
  \label{eq:gamma}
  \end{equation}

Two levels, 0.2 and 0.9, of \texttt{gamma} are used for simulation so that level 0.2 simulates data with low multicollinearity and 0.9 simulates the data with high multicollinearity in \(\mathbf{x}\) respectively.
\item[\textbf{Position of relevant components: (\texttt{relpos})}]
Initial principal components of a non-singular covariance matrix have higher variance than the later ones. If the principal components corresponding to predictors with larger variation are not relevant for a response, this will just increase the noise level in the data. Here we will use two different levels of a position index of predictor components (\texttt{relpos}): a) 1, 2, 3, 4 and b) 5, 6, 7, 8. Predictor components irrelevant for a response make prediction difficult \citep{Helland1994b}. When combined with multicollinearity, this factor can create both easy and difficult cases for both estimation and prediction.
\item[\textbf{Correlation in response variables: (\texttt{eta})}]
Some estimators also use the dependence structure of response for estimation. Here the correlation between the responses is varied through a simulation parameter \texttt{eta} \((\eta)\). The parameter controls the exponential decline of eigenvalues \(\kappa_j, j = 1, \ldots m (\text{ number of responses})\) of \(\boldsymbol{\Sigma_{yy}}\) as,
\begin{equation}
\eta_j = e^{-\kappa(j-1)}, \kappa > 0 \text{ and } j = 1, 2, \ldots m
\label{eq:eta}
\end{equation}

Four levels 0, 0.4, 0.8 and 1.2 of \texttt{eta} are used in the simulations. Level \(\kappa=0\) gives data with uncorrelated response variables, while \(\kappa=1.2\) gives highly correlated response variables.
\end{description}

\begin{figure}
\includegraphics[width=1\linewidth]{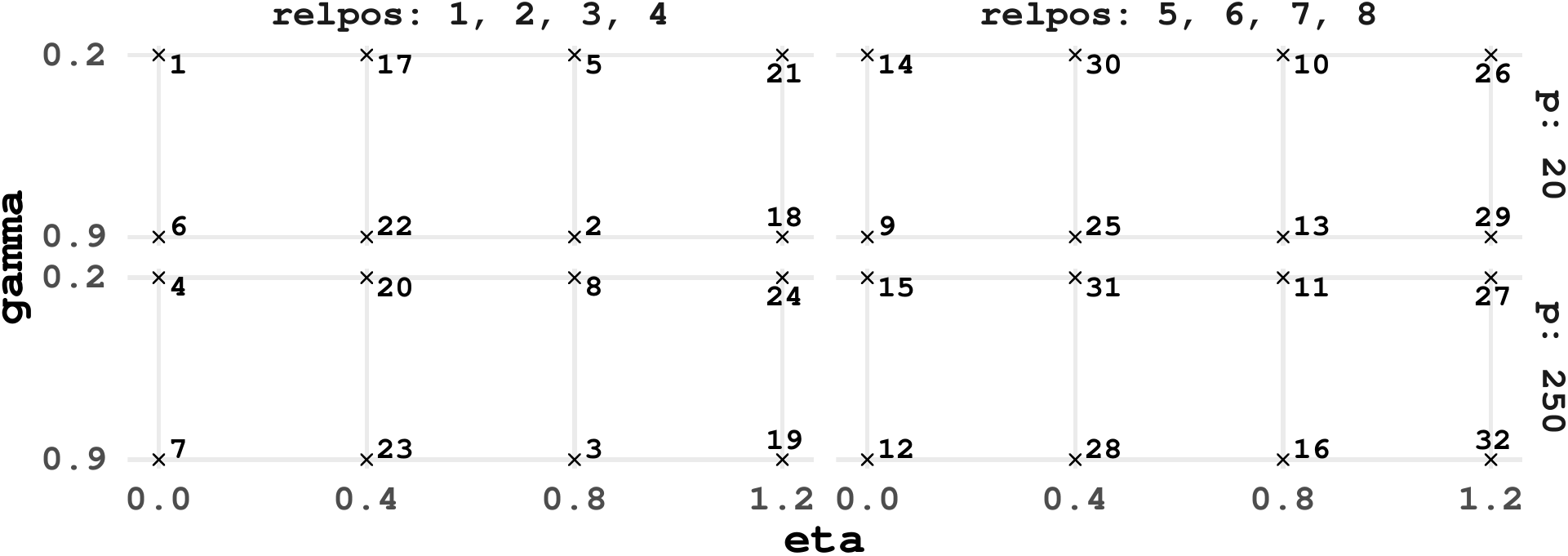} \caption{Experimental Design of simulation parameters. Each point represents an unique data property.}\label{fig:design-plot}
\end{figure}

Here we have assumed that there is only one informative response component. Hence the relevant space of the response matrix has dimension one. In the final dataset all predictors together span the same space as the relevant predictor components and all responses together span the same space as the one informative response component. In addition, the coefficient of determination is fixed at 0.8 for all datasets.

A complete factorial design is adopted using the different levels of factors discussed above to create 32 designs (Figure \ref{fig:design-plot}), each of which gives datasets with unique properties. From each of these design and each estimation method, 50 different datasets are simulated so that each of them has the same true population structure. In total, \(5 \times 32 \times 50\) i.e., 8000 datasets are simulated.

\begin{figure}[H]
\includegraphics[width=1\linewidth]{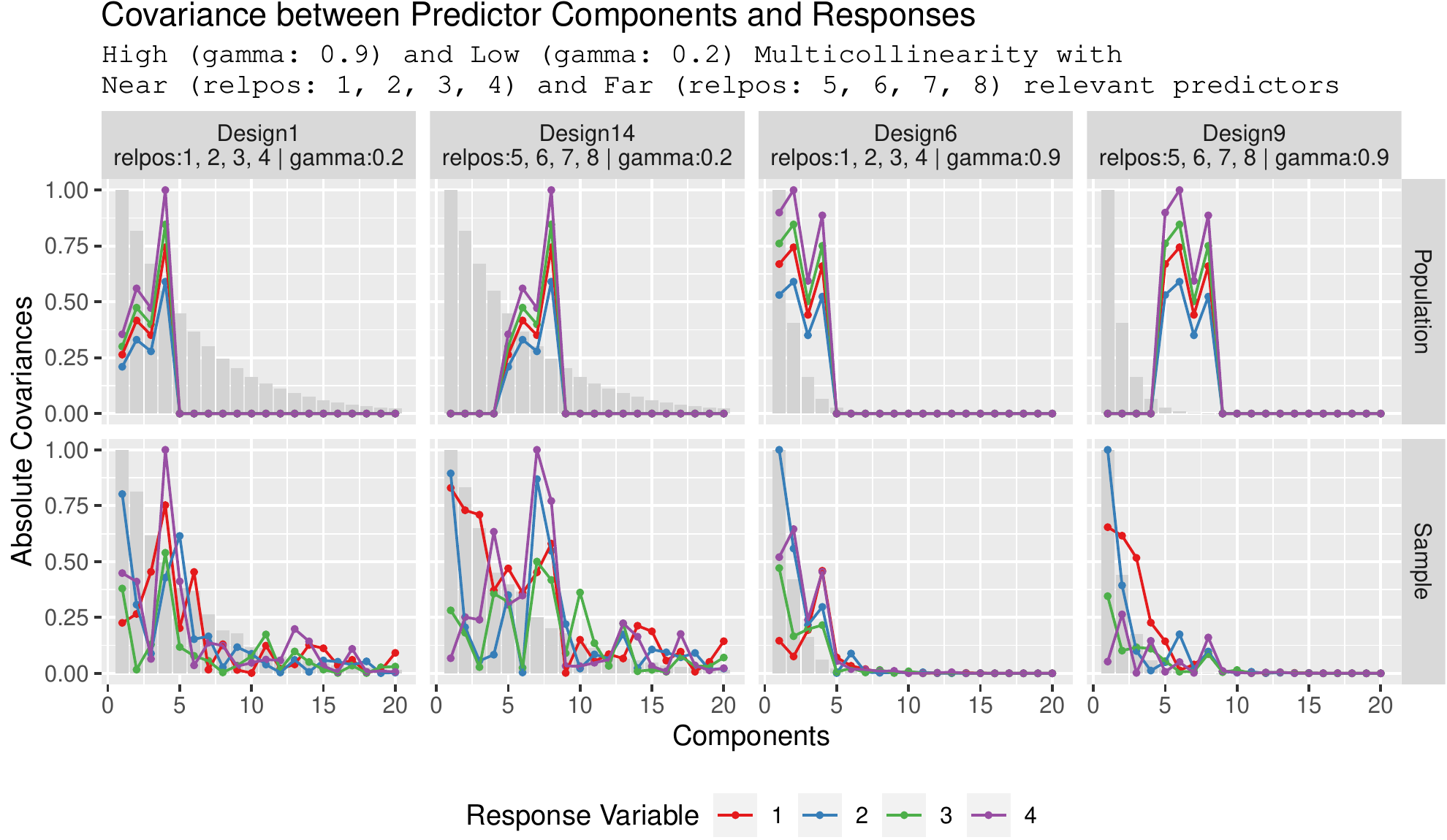} \caption{Covariance between predictor components and each response variable in the population (top), and in the simulated data (bottom) for four different designs. The bars in the background represent the variance of the corresponding components (eigenvalues).}\label{fig:cov-plot}
\end{figure}

The simulation properties are directly reflected in the simulated data. For example, in Figure \ref{fig:cov-plot}, design pairs 1 and 14 as well as 6 and 9 differ in their properties only in terms of position of relevant predictor components, while the design pairs 1 and 6 as well as 9 and 14 differ only in-terms of the level of multicollinearity. The population properties are also reflected in the simulated samples (bottom row Figure \ref{cov-plot}). The combination of these factor levels creates datasets that are easy or difficult with regard to estimation and prediction. We observe from Figure \ref{fig:cov-plot} that it may be difficult to infer the structure of the latent relevant space of \(\mathbf{x}\) from the estimated principal components and their estimated covariances with the observed responses.

\hypertarget{basis-of-comparison}{%
\section{Basis of Comparison}\label{basis-of-comparison}}

The focus of this study is to extend the exploration of \citet{rimal2019pred} to compare the estimation performance of PCR, PLS1, PLS2, Xenv and Senv methods. The performance is measured on the basis of,

\begin{enumerate}
\def\labelenumi{\alph{enumi})}
\tightlist
\item
  average estimation error computed as in \eqref{eq:estimated-est-error}
\item
  the average number of components used by the methods to give minimum estimation error
\end{enumerate}

Let us define the expected estimation error as
\begin{equation}
  \text{MSE}\left(
    \widehat{\boldsymbol{\beta}}
  \right)_{ijkl} =
  \mathsf{E}{\left[
    \frac{1}{\sigma_{y_j}^2}\left(
      \boldsymbol{\beta}_{ij} - \boldsymbol{\widehat{\beta}_{ijkl}}
    \right)^t
    \left(
      \boldsymbol{\beta}_{ij} - \boldsymbol{\widehat{\beta}_{ijkl}}
    \right)
  \right]}
\label{eq:est-error}
\end{equation}
for response \(j = 1, \ldots 4\) in a given design \(i=1, 2, \ldots 32\) and method \(k=1(PCR), \ldots 5(Senv)\) using \(l=0, \ldots 10\) number of components. Here \(\sigma_{y_j}^2\) is the variance of response \(j\). Since both the expectation and the variance of \(\widehat{\boldsymbol{\beta}}\) are unknown, the estimation error is estimated using data from 50 replications as follows,

\begin{equation}
\widehat{\text{MSE}\left(\widehat{\boldsymbol{\beta}}\right)_{ijkl}} =
  \frac{1}{50}\sum_{r=1}^{50}{\left[
    \widehat{\text{MSE}_\circ\left(\widehat{\boldsymbol{\beta}}\right)_{ijklr}}  
  \right]}
\label{eq:estimated-est-error}
\end{equation}
where, \(\widehat{\text{MSE}\left(\widehat{\boldsymbol{\beta}}\right)_{ijkl}}\) is the estimated prediction error averaged over \(r=50\) replicates and,
\[\widehat{\text{MSE}_\circ\left(\boldsymbol{\widehat{\beta}}\right)_{ijklr}} = 
  \frac{1}{\sigma_{y_j}^2}\left[\left(\boldsymbol{\beta}_{ij} -\boldsymbol{\widehat{\beta}_{ijklr}}\right)^t\left(\boldsymbol{\beta}_{ij} - \boldsymbol{\widehat{\beta}_{ijklr}}\right)
\right]\]

Our further discussion revolves around what we will refer to as the \emph{Error Dataset} and the \emph{Component Dataset,} as in the prediction comparison paper \citet{rimal2019pred}. For a given estimation method, design, and response, the component that gives the minimum estimation error averaged over all replicates is selected as,
\begin{equation}
  l_\circ = \operatorname*{argmin}_{l}\left[\frac{1}{50}\sum_{r=1}^{50}{\widehat{\text{MSE}_\circ\left(\widehat{\boldsymbol{\beta}}\right)}_{r}}\right]
  \label{eq:min-err}
\end{equation}
Here we have skipped further indices on \(\boldsymbol{\widehat{\boldsymbol{\beta}}}\) for brevity. The estimation error \(\widehat{\text{MSE}_\circ\left(\widehat{\boldsymbol{\beta}}\right)}\) for every method, design and response corresponding to component \(l_\circ\), computed as \eqref{eq:min-err}, is then regarded as the \emph{error dataset} in the subsequent analysis. Let \(\mathbf{u}_{8000\times4}=(u_j)\), where \(u_j\) is the \(j^\text{th}\) column of \(\mathbf{u}\) denoting the estimation error corresponding to response \(j=1, \ldots 4\) in the context of this dataset. Further, let the number of components that
result in minimum estimation error in each replication and computed as \eqref{eq:min-comp}, comprise the \emph{component dataset}. Let \(\mathbf{v}_{8000\times4}=(v_j)\) where \(v_j\) is the \(j^\text{th}\) column of \(\mathbf{v}\) denoting the outcome variable measuring the number of components used to obtain minimum estimation error corresponding to response \(j=1, \ldots 4\).

\begin{equation}
  l_{\circ} = \operatorname*{argmin}_{l}\left[\widehat{\text{MSE}_\circ\left(\widehat{\boldsymbol{\beta}}\right)}\right]
  \label{eq:min-comp}
\end{equation}

\hypertarget{exploration}{%
\section{Exploration}\label{exploration}}

\begin{figure}[!htb]
\includegraphics[width=1\linewidth]{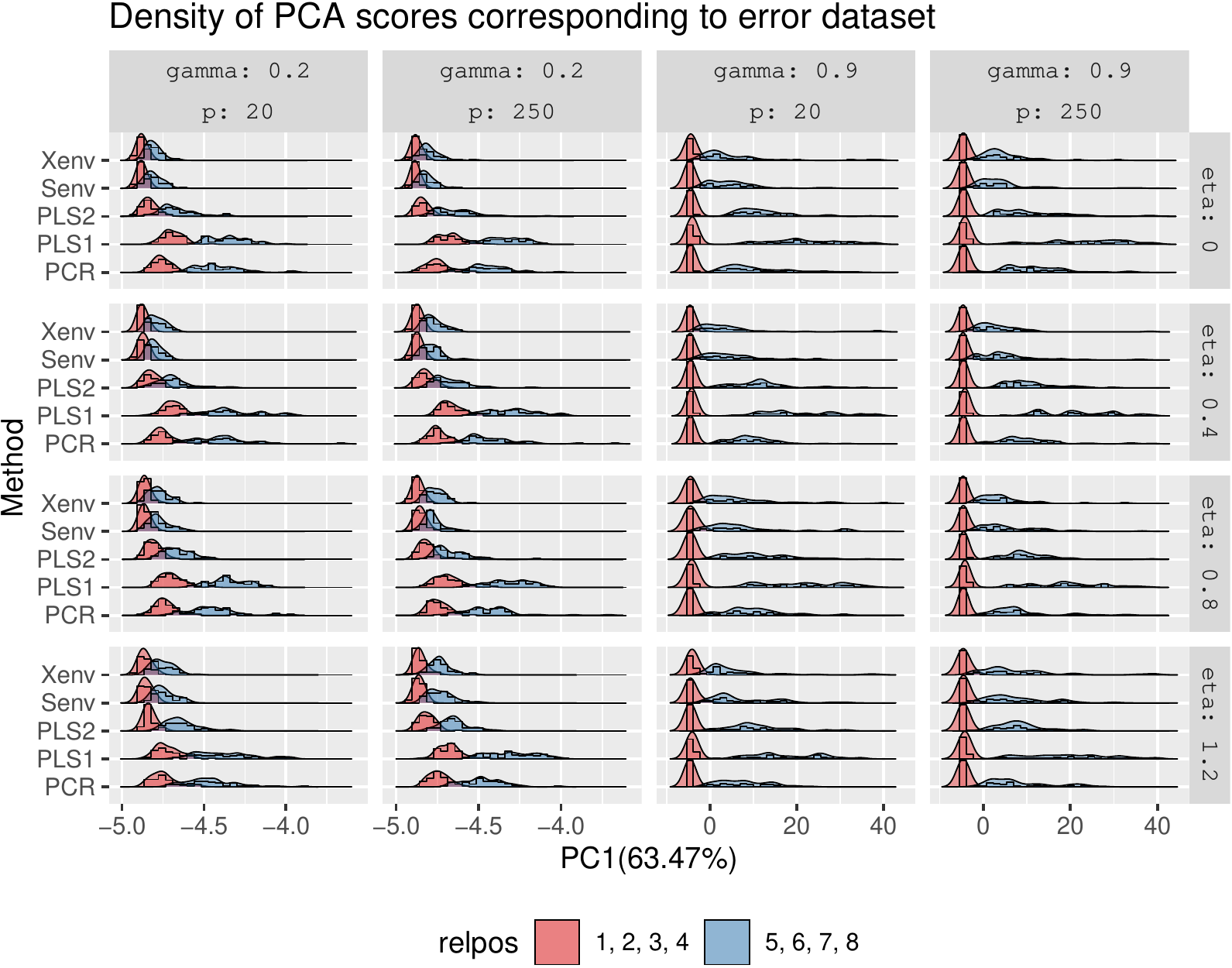} \caption{Scores density corresponding to first principal component of \emph{error dataset} (\(\mathbf{u}\)) subdivided by \texttt{methods}, \texttt{gamma} and \texttt{eta} and grouped by \texttt{relpos}.}\label{fig:est-pca-hist-mthd-gamma-relpos}
\end{figure}

\begin{figure}[!htb]
\includegraphics[width=1\linewidth]{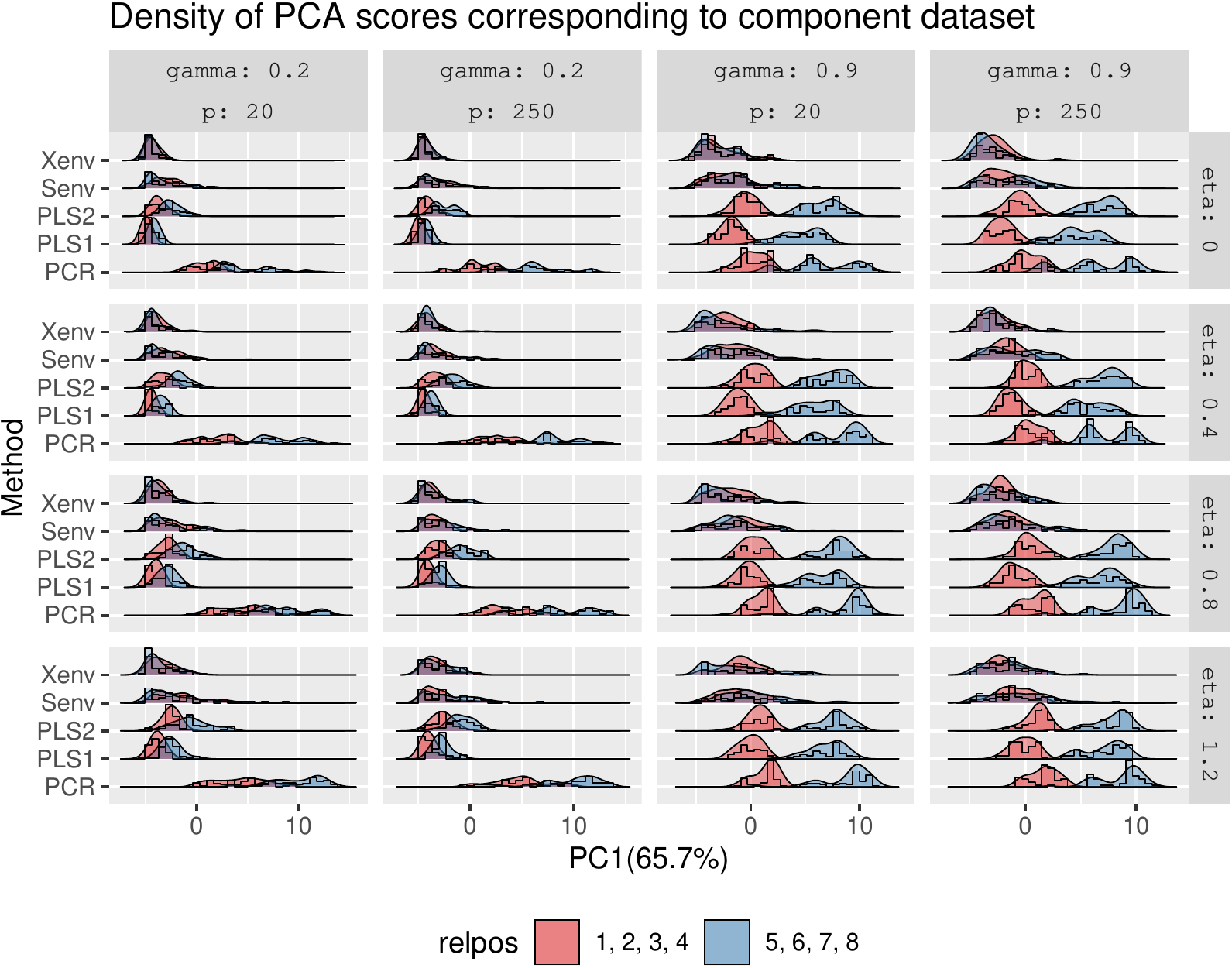} \caption{Score density corresponding to the first principal component of \emph{component dataset} (\(\mathbf{v}\)) subdivided by \texttt{methods}, \texttt{gamma} and \texttt{eta} and grouped by \texttt{relpos}.}\label{fig:comp-pca-hist-mthd-gamma-relpos}
\end{figure}

In this section we explore the variation in the \emph{error dataset} and the \emph{component dataset} by means of Principal Component Analysis (PCA). Let \(\mathbf{t}_u\) and \(\mathbf{t}_v\) be matrices holding the column vectors of the principal component scores corresponding to the \(\mathbf{u}\) and \(\mathbf{v}\) matrices, respectively. The density of the scores in Figure \ref{fig:est-pca-hist-mthd-gamma-relpos} and Figure \ref{fig:comp-pca-hist-mthd-gamma-relpos} correspond to the first principal component of \(\mathbf{u}\) and \(\mathbf{v}\), i.e.~the first column of \(\mathbf{t}_u\) and \(\mathbf{t}_v\) respectively. Here higher scores correspond to larger estimation error and vice versa.

Figure \ref{fig:est-pca-hist-mthd-gamma-relpos} shows a clear difference in the effect of low and high multicollinearity on estimation error. In the case of low multicollinearity (\texttt{gamma:\ 0.2}), the estimation errors are in general smaller and have lesser variation compared to high multicollinearity (\texttt{gamma:\ 0.9}). In particular we observe that the envelope methods have small estimation errors in the low multicollinearity cases compared to the other methods. On the other hand, the envelope methods tend to have increased estimation error in cases of highly correlated responses (\texttt{eta:\ 1.2}), whereas there is no effect of this correlation in other methods.

Furthermore, position of the relevant predictor components has a noticeable effect on estimation error for all methods. When relevant predictors are at position 5, 6, 7, 8, the components at positions 1, 2, 3, 4, which carry most of the variation, become irrelevant. These irrelevant components with large variation add noise to the model and consequently increases the estimation error. The effect intensifies with highly collinear predictors (\texttt{gamma}=0.9). Designs with high multicollinearity and relevant predictors at position 5, 6, 7, 8 are relatively difficult to model for all the methods. Although these difficult designs have a large effect on estimation error, their effect on prediction error is less influential \citep{rimal2019pred}.

In the case of the \emph{component dataset} (Figure \ref{fig:comp-pca-hist-mthd-gamma-relpos}), PCR, PLS1 and PLS2 methods have in general used a larger number of components in the case of high multicollinearity compared to low. Surprisingly, the envelope methods (Senv and Xenv) have mostly used a distinctly smaller number of components in both cases of multicollinearity compared to other methods.

The plot also shows that there is no clear effect of the correlation between response variables (\texttt{eta}) on the number of components used to obtain minimum estimation error.

A clear interaction between the position of relevant predictors and the multicollinearity, which is visible in the plot, suggests that the methods use a larger number of components when the relevant components are at position 5, 6, 7, 8. Additionally, the use of components escalate and the difference between the two levels of \texttt{relpos} becomes wider in the case of high multicollinearity in the predictor variables. Such performance is also seen the case of prediction error (See \citet{rimal2019pred}), however, the number of components used for optimization of prediction is smaller than in the case of estimation. Even when the relevant components are at position 5, 6, 7, 8, the envelope methods, in contrast to other methods, have used an almost similar number of components as in the case of relevant components at position 1, 2, 3, 4. This shows that the envelope methods identify the predictor space relevant to the response differently, from the other methods and with very few numbers of latent components. This is particularly the case when multicollinearity in \(\mathbf{x}\) is high.

The following sub-section explores in particular the prediction and estimation errors and the estimated regression coefficient of Simultaneous Envelope and Partial Least Squares for a design having high multicollinearity, and with predictor components at positions 5, 6, 7, 8. Here we will use the design with \(n>p\) and two levels of correlation between the responses. These correspond to Design-9 and Design-29 in our simulations.

\begin{figure}
\includegraphics[width=1\linewidth]{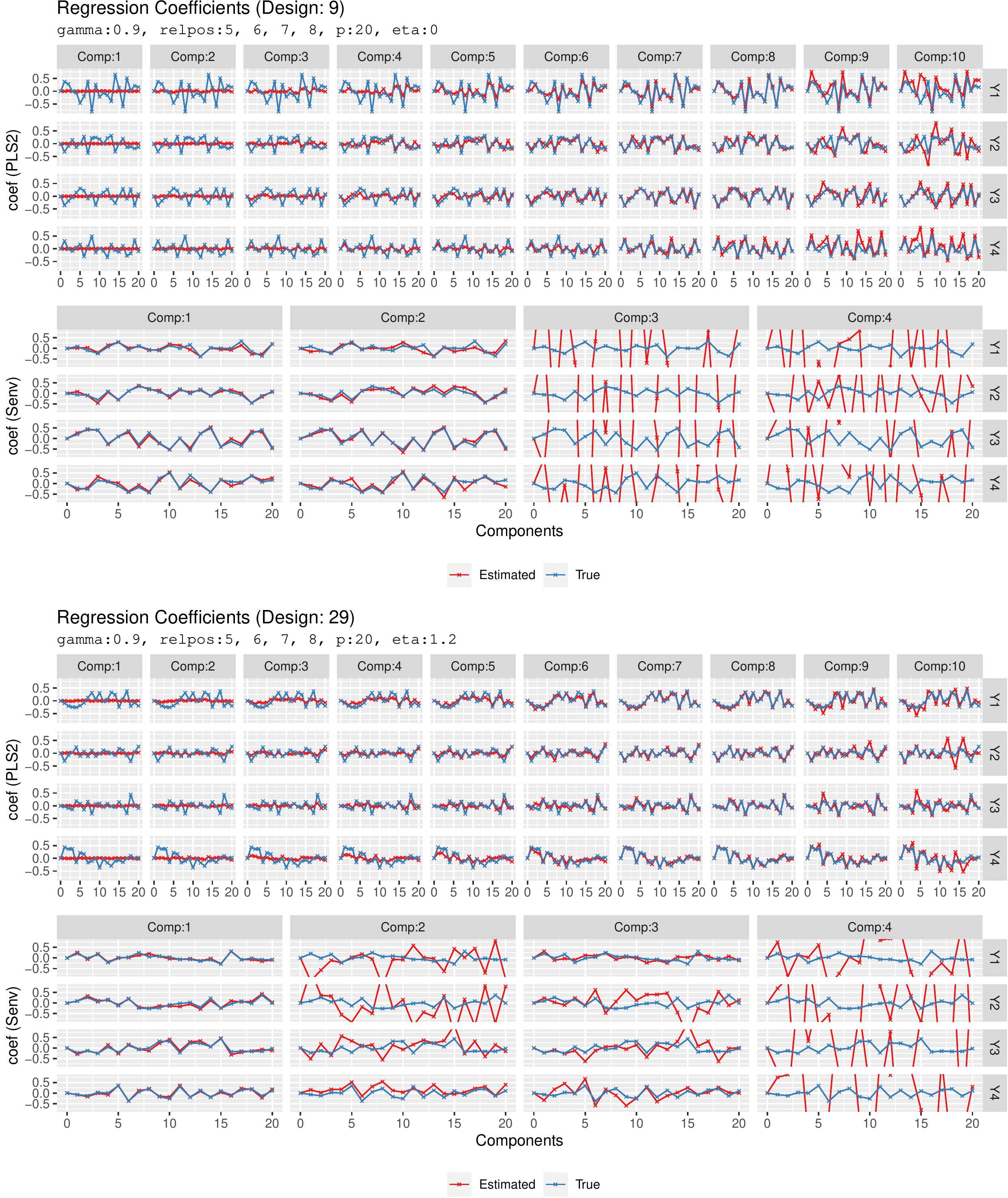} \caption{Regression Coefficients (coef) estimated by PLS2 and Simultaneous Envelope methods on the data based on Design 9 and 29.}\label{fig:coef-plot}
\end{figure}

\begin{figure}
\includegraphics[width=1\linewidth]{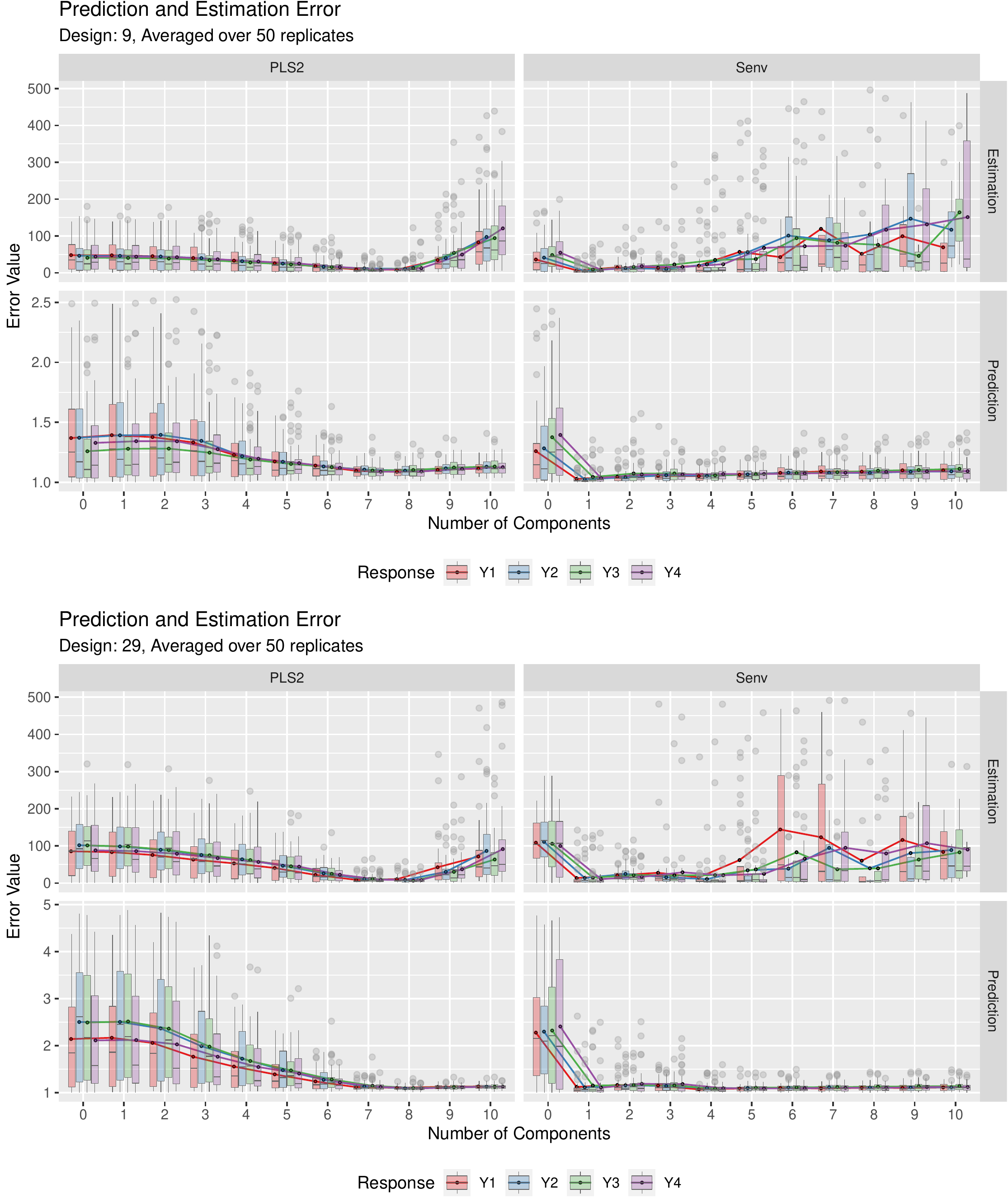} \caption{Minimum prediction and estimation error for PLS2 and Simultaneous Envelope methods. The point and lines are averaged over 50 replications.}\label{fig:err-plot}
\end{figure}

Figure \ref{fig:err-plot} shows a clear distinction between the modelling approach of PLS2 and Senv methods for the same model based on Design 9 (top) and Design 29 (bottom). In both of the designs, PLS2 has both minimum prediction error and minimum estimation error obtained using seven to eight components and the estimated regression coefficients approximate the true coefficients. In contrast, the Senv method has approached the minimum prediction and minimum estimation error using one to two components and the corresponding estimated regression coefficients approximate the true coefficients (Figure \ref{fig:coef-plot}). Despite having contrasted modelling results for a dataset with similar properties, the minimum errors produced by them are comparable in the case of Design 9 (See Table \ref{tab:min-err-dgn9}). However, in the case of Design 29, estimation error corresponding to PLS1 and envelope methods are much higher than PCR and PLS2. It is interesting to see that despite having large estimation error, the prediction error corresponding to the envelope methods are much smaller in this design.

Here the response dimension for the simultaneous envelope has been fixed at two components, which might have affected its performance, however, both envelope methods had performed much better with the same restriction in the case of prediction.

Figure \ref{fig:err-plot} also shows in both designs that Senv has large estimation errors when the number of components is not optimal. This is also true for the PLS2 model, however, the extent of this variation is noticeably large for the Senv method. A similar observation as Senv is also found in Xenv method while PCR and PLS1 are closer to the PLS2 in terms of their use of components in order to produce the minimum error (See Table \ref{tab:min-err-dgn9}).

In addition to the prediction and estimation error, Figure \ref{fig:coef-plot} gives a closer view of how the average coefficients corresponding to these methods approximate to the true values. Here PLS2 has used seven to eight components to reach the closest approximation to the true coefficients, but with increasing errors after including more components than eight. This departure from true coefficients is usual for PLS when the relevant components are at 1, 2, 3, 4 whereas PCR has shown more stable result in such situations. Further, the envelope methods have presented their ability to converge estimates to the true value in just one or two components. However, one should be cautious about determining the optimal components in these methods due to a highly unstable and large error in non-optimal cases.

\begin{table}[t]

\caption{\label{tab:min-err-dgn9}Minimum Prediction and Estimation Error for Design 9}
\centering
\begin{tabu} to \linewidth {>{\raggedleft}X>{\raggedleft}X>{\raggedleft}X>{\raggedleft}X>{\bfseries\em}r>{\bfseries\em}r>{\raggedleft}X}
\toprule
Design & Response & PCR & PLS1 & PLS2 & Senv & Xenv\\
\midrule
\addlinespace[0.3em]
\multicolumn{7}{l}{\textbf{Design 9}}\\
\addlinespace[0.3em]
\multicolumn{7}{l}{\textbf{Estimation Error}}\\
\hspace{1em}\hspace{1em}9 & 1 & 8.56 (8) & 13.23 (6) & 8.17 (8) & 6.65 (1) & 5.73 (1)\\
\hspace{1em}\hspace{1em}9 & 2 & 7.94 (8) & 14.42 (6) & 10.65 (8) & 5.06 (1) & 5.35 (1)\\
\hspace{1em}\hspace{1em}9 & 3 & 7.02 (8) & 15.9 (6) & 8.22 (7) & 8.55 (1) & 5 (1)\\
\hspace{1em}\hspace{1em}9 & 4 & 9.26 (8) & 13.14 (7) & 8.29 (7) & 8.19 (1) & 4.78 (1)\\
\addlinespace[0.3em]
\multicolumn{7}{l}{\textbf{Prediction Error}}\\
\hspace{1em}\hspace{1em}9 & 1 & 1.08 (8) & 1.1 (7) & 1.09 (8) & 1.03 (1) & 1.03 (1)\\
\hspace{1em}\hspace{1em}9 & 2 & 1.09 (8) & 1.11 (7) & 1.1 (8) & 1.03 (1) & 1.03 (1)\\
\hspace{1em}\hspace{1em}9 & 3 & 1.08 (8) & 1.1 (7) & 1.1 (7) & 1.04 (1) & 1.03 (1)\\
\hspace{1em}\hspace{1em}9 & 4 & 1.09 (8) & 1.1 (7) & 1.09 (7) & 1.04 (1) & 1.03 (1)\\
\addlinespace[0.3em]
\multicolumn{7}{l}{\textbf{Design 29}}\\
\addlinespace[0.3em]
\multicolumn{7}{l}{\textbf{Estimation Error}}\\
\hspace{1em}\hspace{1em}29 & 1 & 6.16 (8) & 13.64 (7) & 8.67 (7) & 13.45 (1) & 13.05 (1)\\
\hspace{1em}\hspace{1em}29 & 2 & 6.29 (8) & 12.3 (7) & 8.49 (8) & 13.62 (1) & 10.98 (1)\\
\hspace{1em}\hspace{1em}29 & 3 & 6.73 (8) & 13.03 (7) & 6.54 (8) & 14.72 (1) & 16.24 (1)\\
\hspace{1em}\hspace{1em}29 & 4 & 6.28 (8) & 12.51 (7) & 8.66 (8) & 10.76 (1) & 10.27 (1)\\
\addlinespace[0.3em]
\multicolumn{7}{l}{\textbf{Prediction Error}}\\
\hspace{1em}\hspace{1em}29 & 1 & 1.09 (8) & 1.1 (8) & 1.1 (8) & 1.07 (4) & 1.1 (5)\\
\hspace{1em}\hspace{1em}29 & 2 & 1.1 (8) & 1.11 (8) & 1.09 (8) & 1.1 (5) & 1.11 (1)\\
\hspace{1em}\hspace{1em}29 & 3 & 1.1 (8) & 1.1 (8) & 1.1 (8) & 1.09 (4) & 1.13 (5)\\
\hspace{1em}\hspace{1em}29 & 4 & 1.09 (8) & 1.11 (8) & 1.09 (8) & 1.09 (5) & 1.11 (1)\\
\bottomrule
\end{tabu}
\end{table}

Despite having a large variation in prediction and estimation error, the envelope based methods have produced a better result even for the difficult data cases as shown for Design 9.

\hypertarget{analysis}{%
\section{Analysis}\label{analysis}}

A statistical analysis using a Multivariate Analysis of variance (MANOVA) model is performed on both the \emph{error dataset} and the \emph{component dataset} in order to better understand the association between data properties and the estimation methods. Let the corresponding MANOVA models be termed as the \emph{error model} \eqref{eq:err-model} and the \emph{component model} \eqref{eq:comp-model} in the following. In the MANOVA model, we will consider the interaction of simulation parameters (\texttt{p}, \texttt{gamma}, \texttt{eta}, and \texttt{relpos}) and \texttt{Method} The models are fitted using correspondingly the \emph{error dataset} (\(\mathbf{u}\)) and the \emph{component dataset} (\(\mathbf{v}\)).

\textbf{Error Model:}
\begin{equation}
  \mathbf{u} = \boldsymbol{\mu} +
  (\texttt{p} + \texttt{gamma} + \texttt{eta} +
    \texttt{relpos} + \texttt{Methods})^3 +
    \boldsymbol{\varepsilon}
  \label{eq:err-model}
\end{equation}

\textbf{Component Model:}
\begin{equation}
  \mathbf{v} = \boldsymbol{\mu} +
  (\texttt{p} + \texttt{gamma} + \texttt{eta} +
    \texttt{relpos} + \texttt{Methods})^3 +
    \boldsymbol{\varepsilon}
  \label{eq:comp-model}
\end{equation}

where, \(\mathbf{u}\) corresponds to the estimation errors in \emph{error dataset} and \(\mathbf{v}\) corresponds to the number of components used by a method to obtain minimum estimation error in the \emph{component dataset}.

\begin{figure}[H]
\includegraphics[width=1\linewidth]{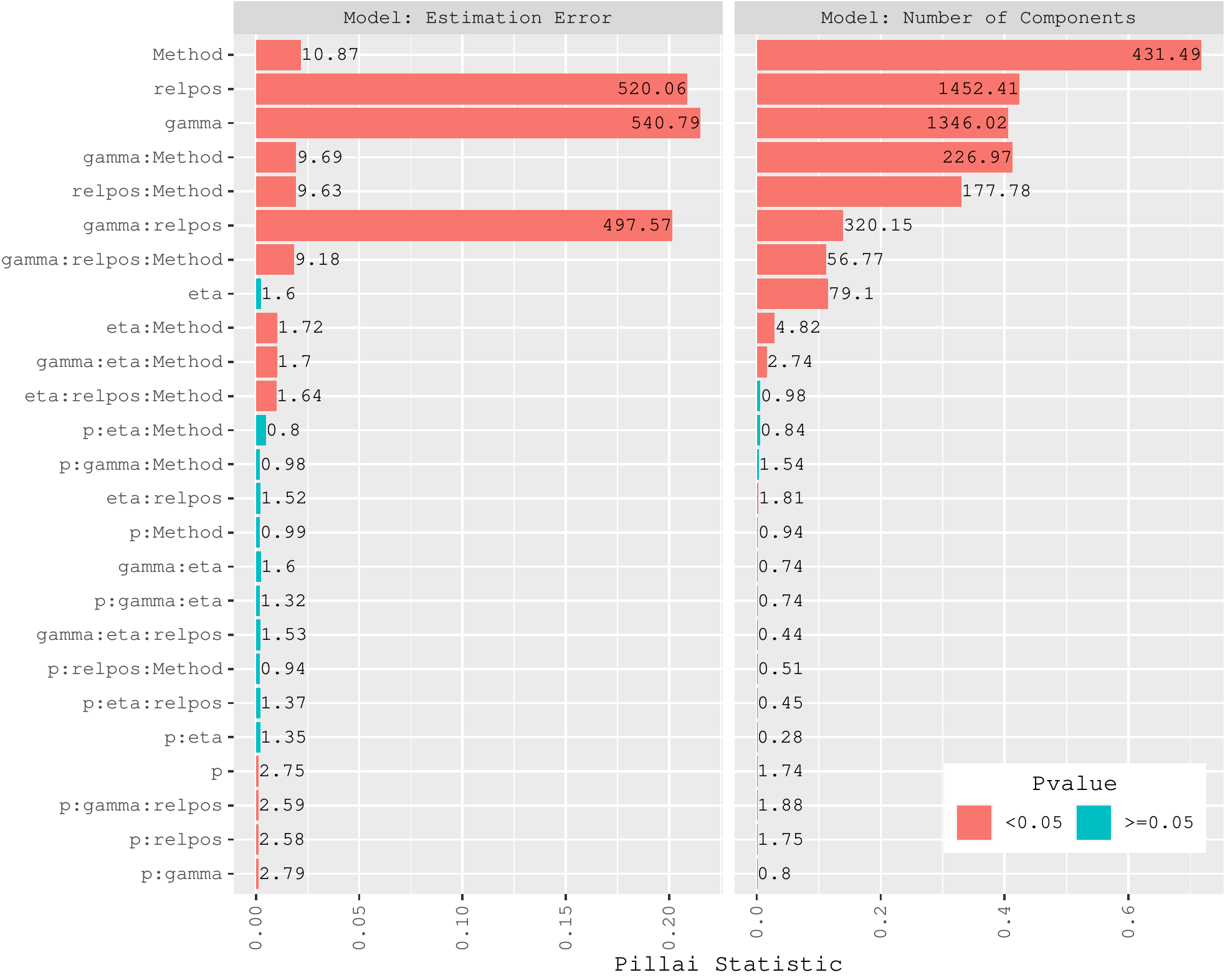} \caption{Pillai Statistic and F-value for the MANOVA model. The bar represents the Pillai Statistic and the text labels are F-value for the corresponding factor.}\label{fig:manova-plot}
\end{figure}

To make the analysis equivalent to \citet{rimal2019pred}, we have also used Pillai's trace statistic for accessing the result of MANOVA. Figure \ref{fig:manova-plot} plots the Pillai's trace statistics as bars with corresponding F-values as text labels. The leftmost plot corresponds to the \emph{error model} and the rightmost plot corresponds to the \emph{component model}. Here we use the custom R-notation indicating interactions up to order three for the parameters within the brackets.

\begin{description}
\tightlist
\item[\textbf{Error Model:}]
Unlike for the prediction error in \citet{rimal2019pred}, \texttt{Method} has a smaller effect, while the amount of multicollinearity, controlled by the \texttt{gamma} parameter, has a larger effect in the case of estimation error (Figure \ref{fig:manova-plot}). In addition, the position of relevant components and its interaction with the \texttt{gamma} parameters also have substantial effects on the estimation error. This also supports the results seen in the \protect\hyperlink{exploration}{Exploration} section where relevant predictors at position 5, 6, 7, 8 combined with high multicollinearity creates a large uninformative variance in the components 1, 2, 3, 4 making the design difficult with regards to estimation. The effect of this on the estimation error is much larger than on the prediction error.

Furthermore, the \texttt{eta} factor controlling the correlation between the responses, and its second-order interaction with other factors except for the number of predictors is significant. The effect is also comparable with the main effect of \texttt{Method} and \texttt{eta}.
\item[\textbf{Component Model:}]
Although \texttt{Method} does not have a large impact on the estimation error, the \emph{component model} in Figure \ref{fig:manova-plot} (right) shows that the methods are significantly different and has a huge effect on the number of components they use to obtain the minimum estimation error. The result also corresponds to the case of prediction error in \citet{rimal2019pred}. However, the F-value corresponding the \texttt{relpos} and \texttt{gamma} shows that the importance of these factors is much stronger compared to the case of prediction error.
\end{description}

The following section will further explore the effects of individual levels of different factors.

\hypertarget{effect-analysis-of-the-error-model}{%
\subsection{Effect Analysis of the Error Model}\label{effect-analysis-of-the-error-model}}

\begin{figure}
\includegraphics[width=1\linewidth]{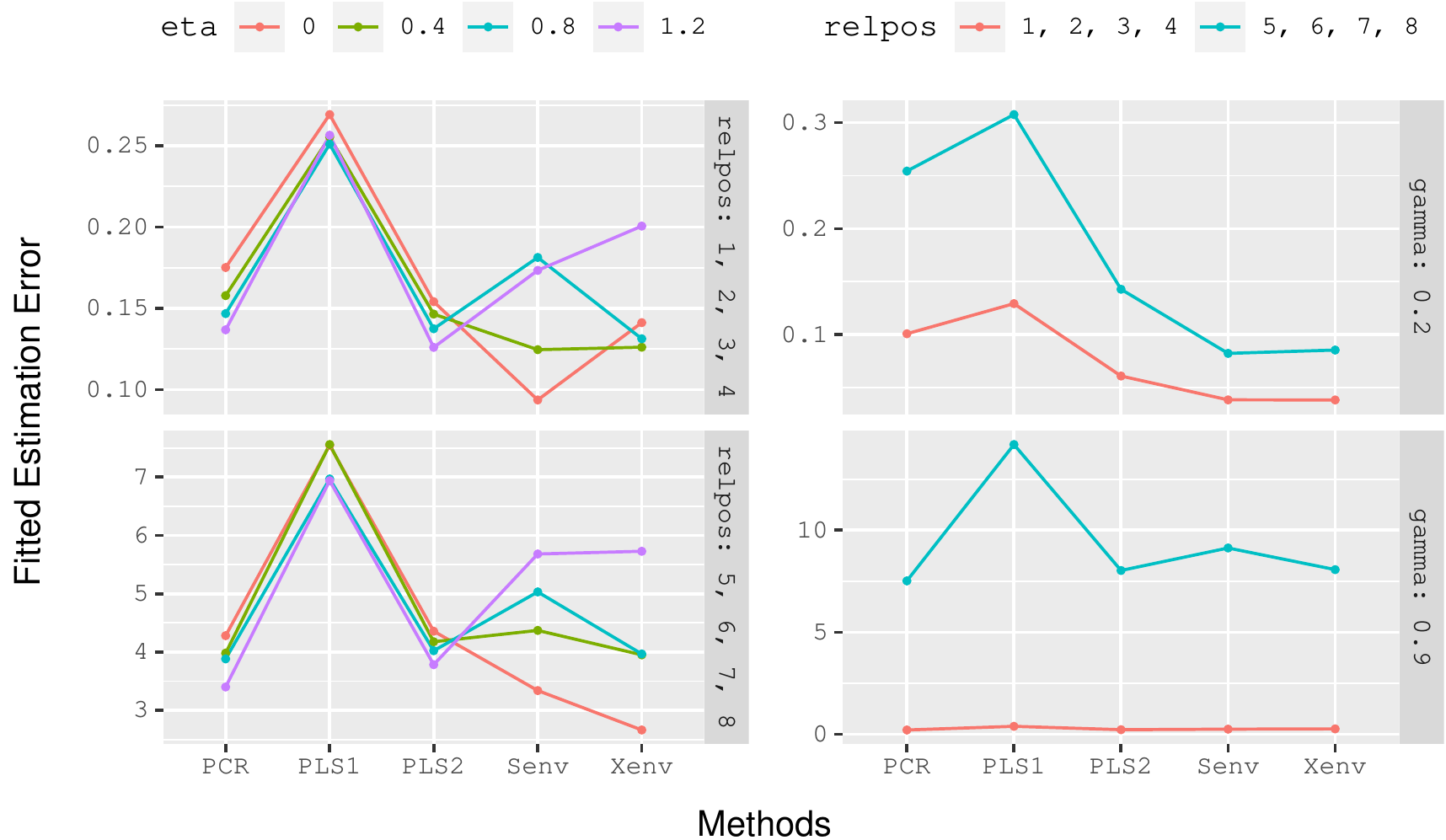} \caption{Effect plot of some interactions of the MANOVA corresponding to fitted \emph{error model}}\label{fig:est-eff-plots}
\end{figure}

In figure \ref{fig:est-eff-plots} (left), the effect of correlation between the responses controlled by the \texttt{eta} parameter has a clear influence on the estimation error for the envelope methods. In the case of designs with uncorrelated responses, envelope methods have on average smallest estimation errors. While PCR and PLS2, being somewhat invariant to the effect of this correlation structure, have performed better than the envelope methods in the designs with highly correlated responses.

For all methods, the error in the case of relevant predictors at positions 5, 6, 7, 8 is huge as compared to the case where relevant predictors are at positions 1, 2, 3, 4.

Figure \ref{fig:est-eff-plots} (right) shows a large difference in the effect of the two levels of the position of relevant components, especially in the designs with high multicollinearity. In the case of high multicollinearity, all methods have noticeable poorer performance compared to the case of low multicollinearity.

Finally, we note that the average estimation error corresponding to envelop methods in the designs with low multicollinearity is smaller than for the other methods.

\hypertarget{effect-analysis-of-the-component-model}{%
\subsection{Effect Analysis of the Component Model}\label{effect-analysis-of-the-component-model}}

\begin{figure}[!htb]
\includegraphics[width=1\linewidth]{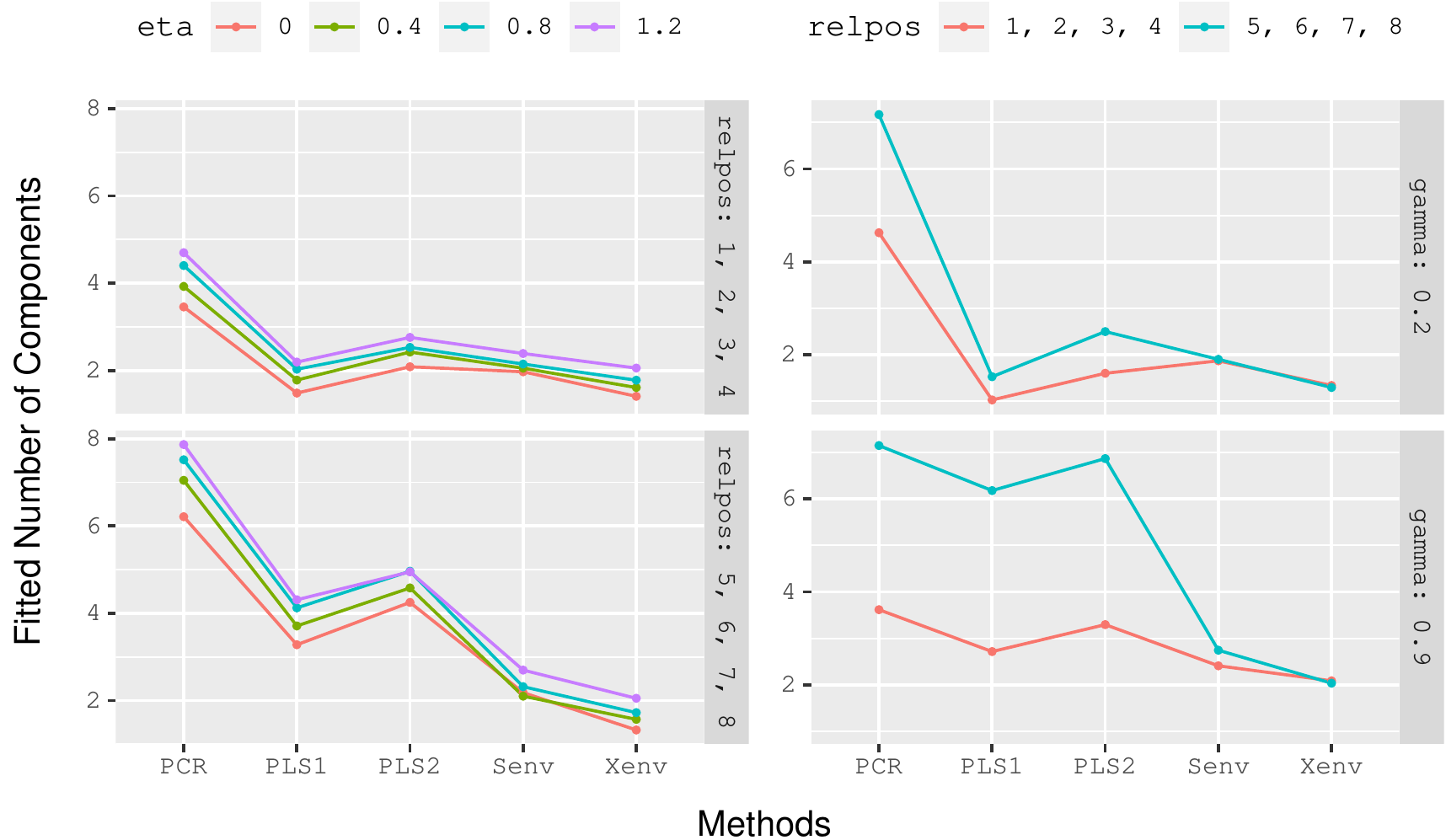} \caption{Effect plots of some interactions of the multivariate linear model corresponding to the \emph{component model}.}\label{fig:comp-eff-plots}
\end{figure}

In the case of the fitted \emph{component model}, envelope methods are the clear winner in almost all designs. In the case of low multicollinearity and position of relevant predictors at 1, 2, 3, 4, PLS1 has obtained the minimum estimation error similar to the envelope methods, however, in the case of high multicollinearity PLS1 has also used a fairly large number of components to obtain the minimum estimation error. Although the envelope methods have comparable minimum estimation error in some of the designs, in almost all the designs these methods have used 1-2 components on average. The effect of the correlation in the response has minimal effect on the number of components used by the methods. The design 9, which we have considered in the previous section, has minimum estimation error for both envelope methods using only one predictor component. In design 29, where the envelope methods have poorer performance than the other methods due to highly correlated responses, the number of components used by them is still one. This corresponds to the results seen in Figure \ref{fig:comp-eff-plots}. As seen previously, PCR uses, in general, a larger number of components than the other methods.

\hypertarget{discussion-and-conclusion}{%
\section{Discussion and Conclusion}\label{discussion-and-conclusion}}

The overall performance of all methods highly depends on the nature of the data. The MANOVA plots show that most of the simulation parameters, except \texttt{p}, has significant interaction with the methods. In addition, the high interaction of \texttt{gamma} with the \texttt{relpos} parameter suggests to carefully consider the number of relevant predictor components in the case of highly multicollinear data since this choice may have a large effect on the results. Although the interaction does not have this extent of influence in prediction, one should be careful about interpreting the estimates. In such cases, careful validation of model complexity, preferably using cross-validation or test data is advisable also for estimation purposes.

Designs with low multicollinearity and independent responses are in favour of envelope methods. The methods have produced the smallest prediction and estimation error with significantly few numbers of components in these designs. However, as the correlation in the responses increases, the estimation error in envelope methods in most cases also increases noticeably. This indicates that the reduction of the response space becomes unstable with high collinearity between the responses for the envelope methods. Despite the interaction of the \texttt{eta} parameter with the method is significant, the extent of its effect is rather small compared to both main and interaction effect of \texttt{gamma} and \texttt{relpos}.

The effect of the number of variables is negligible in all cases for all designs. Here the use of principal components for reducing the dimension of \(n<p\) designs, as in \citet{rimal2019pred}, has been useful so that we were able to model the data using envelope methods without losing too-much variation in the data.

Both prediction and estimation corresponding to PCR methods are found to be stable even when the non-optimal number of components are used. The PLS1 method, which models the responses separately, is in general performing poorer than other methods. Unlike in prediction comparison, the performance of the envelope methods is comparable to the others except for the use of the number of components to obtain the minimum estimation error. The envelope methods have used 1-2 components in almost all designs, which is quite impressive. However, non-optimal number of components can lead to large estimation error, so one should be careful in this respect while using the envelope methods. Both PLS1 and PLS2 use a smaller number of components when the relevant components are at positions 1, 2, 3, 4. However, both methods used 7-8 components for the designs with relevant components at positions 5, 6, 7, 8.

We expect the results from this study may help researchers, working on theory, application and modelling, to understand these methods and their performance on data with varying properties.

The first part of this study \citep{rimal2019pred} on prediction comparison should be considered to obtain a comprehensive view of this comparison. A shiny \citep{shiny} web application at \url{http://therimalaya.shinyapps.io/Comparison} allows readers to explore all the visualizations for both prediction and estimation comparisons. In addition, a GitHub repository at \url{https://github.com/therimalaya/04-estimation-comparison} can be used to reproduce this study.

\addcontentsline{toc}{section}{References}

\hypertarget{refs}{}

\renewcommand\refname{References}
\bibliography{References.bib}

\end{document}